\documentclass[prd,floatfix,preprintnumbers,letterpaper,twocolumn]{revtex4}
\usepackage{amsmath,amssymb,graphicx}
\usepackage{hyperref}

\begin{document}
\title{Cosmic opacity: cosmological-model-independent tests from gravitational waves and Type Ia Supernova}

\author{Jing-Zhao Qi$^{1}$}
\author{Shuo Cao$^{2\ast}$}
\author{Yu Pan$^{3}$}
\author{Jin Li$^{4}$}

\affiliation{$^1$ Department of Physics, College of Sciences, Northeastern University, Shenyang 110819, China;\\
$^2$ Department of Astronomy, Beijing Normal University, Beijing,
100875, China; \emph{caoshuo@bnu.edu.cn}; \\
$^3$ College of Science, Chongqing University of Posts and
Telecommunications, Chongqing 400065, China;\\
$^4$ Department of Physics, Chongqing University, Chongqing 400030,
China}

\begin{abstract}
In this paper, we present a scheme to investigate the opacity of the
Universe in a cosmological-model-independent way, with the
combination of current and future available data in gravitational
wave (GW) and electromagnetic (EM) domain. In the FLRW metric, GWs propagate freely
through a perfect fluid without any absorption and dissipation,
which provides a distance measurement unaffected by the cosmic
opacity. Focusing on the simulated
data of gravitational waves from the third-generation gravitational
wave detector (the Einstein Telescope, ET), as well as the
newly-compiled SNe Ia data (JLA and Pantheon sample), we find an
almost transparent universe is strongly favored at much higher
redshifts ($z\sim 2.26$). Our results suggest that, although the
tests of cosmic opacity are not significantly sensitive to its
parametrization, a strong degeneracy between the cosmic opacity
parameter and the absolute \textit{B}-band magnitude of SNe Ia is
revealed in this analysis. More importantly, we obtain that future
measurements of the luminosity distances of gravitational waves
sources will be much more competitive than the current analyses,
which makes it expectable more vigorous and convincing constraints
on the cosmic opacity (and consequently on background physical
mechanisms) and a deeper understanding of the intrinsic properties of
type Ia supernovae in a cosmological-model-independent way.
\end{abstract}

\maketitle

\section{Introduction} \label{introduction}

One of the most important issues of the modern cosmology lies in the
discovery that our universe is undergoing an accelerated expansion
at the present stage, through the observations of unexpected dimming
of type Ia supernovae (SNe Ia) \citep{riess1998supernova}. In the
framework of general relativity (GR), a mysterious substance with
negative pressure, dubbed as dark energy, was proposed to explain
this acceleration
\citep{ratra1988cosmological,caldwell2002phantom,zhu2004generalized,cao2010mn,cao2011prd,cao2011aa}.
However, another mechanism attempts to explain this observed SNe Ia
dimming, i.e., whether the light intensity of a supernova is
diminished because of the photon absorption or scattering of dust in
Milky Way, intervening galaxies or the host galaxy
\citep{csaki2002dimming}. Although subsequent observations such as
large-scale structure \citep{tegmark2004cosmological}, baryon
acoustic oscillation (BAO) \citep{eisenstein2005detection} and
cosmic microwave background (CMB) anisotropy \citep{spergel2003wmap}
independently confirmed the accelerating expansion of the Universe,
the question of whether the universe is transparent still needs to
be confronted, as the acceleration rate and the cosmological
parameters determined by SNe Ia observations are highly dependent on
the dimming effect. For instance, a recent analysis
\citep{nielsen2016marginal} seems to imply only a marginal evidence
for this widely accepted claim, if rigorous statistical tests are
performed on these standardizable candles with the varying shape of
the light curve and extinction by dust. This motivates the need to
probe other plausible mechanisms for this observed SNe Ia dimming.

The general methodology of testing the cosmic opacity focuses on the
distance duality relation (DDR), which connects the luminosity
distance $D_L$ and angular diameter distance (ADD) $D_A$ at the same
redshift, $\frac{D_L(z)}{D_A(z)}\left(1+z\right)^{-2} = 1$. Having
been derived from the reciprocity law, the DDR holds in whatever
cosmology provided the space-time is Riemannian and that the number
of photons is conserved. The former condition, which is related to
the foundations of the gravity theory, could be used to probe the
possible existence of exotic physics in the theory of gravity.
Meanwhile, if one can take it for granted, a more interesting
possibility is to test whether there are sources of attenuation
(like gray dust) or brightening (as gravitational lensing) along the
light path \citep{cao2011testing}.

From the observational point of view, the measurement of the
luminosity distance will be affected when the Universe is opaque. In
recent works, there are many papers
\citep{li2013cosmic,nair2012cosmic,more2009cosmic,chen2012cosmic,cao2016gas,jesus2017testing}
devoted to investigating the cosmic opacity under the assumption
that the violation of DDR is generated by the non-conservation of
the photon number, in which type Ia supernovae are the ideal tool to
estimate the luminosity distances, while the angular diameter
distances are derived from various astrophysical probes. Although
ADDs are much more difficult to measure, some significant steps
forward have been progressed recently based on the
Sunyaev-Zel'dovich effect together with X-ray emission of galaxy
clusters, estimates of the cosmic expansion $H(z)$ from cosmic
chronometers, measurements of the gas mass fraction in galaxy
clusters and observations of strong gravitational lensing systems.
Reference \citep{li2013cosmic} made a joint ADD analysis with two
galaxy cluster samples compiled by
\cite{de2005measuring,bonamente2006determination} and performed
cosmological-model-independent tests for the cosmic opacity. The
final results showed that a transparent universe is ruled out by the
\citet{bonamente2006determination} sample at $68.3\%$ confidence
level (C.L.), which demonstrated the importance of considering the
dimming effect of SNe Ia, given the compatibility of results derived
by using angular diameter distances and luminosity distances,
respectively \citep{cao2014cosmic}. Further papers
\citep{nair2012cosmic,more2009cosmic,chen2012cosmic} have also
noticed this disagreement in cosmographic studies using BAO as a
source of angular diameter distances. More recently, some
substantial progress has been made in the measurements of the Hubble
parameter $H(z)$, which are combined with different sub-samples of
SNe Ia observations to quantify the cosmic opacity
\citep{holanda2013model,liao2013testing}. However, it is worth
noting that $H(z)$ describes the expansion rate of the universe
rather than the distance, i.e., the angular diameter distance
obtained by integrating these scattered points will inevitably lead
to large uncertainties, which indicated the importance of taking the
correlations between different redshifts into account
\citep{liao2015universe}. More importantly, considering the limited
sample size of $H(z)$ measurements, one has also to take care of the
errors due to the mismatch between the $H(z)$ redshift and the
closest SNe Ia in the companion SNe Ia sample adopted. The
cosmological constraining power of these ADD measurements, derived
in the electromagnetic (EM) domain, could be significantly affected
by large observational uncertainties.

An alternative opacity-free distance indicator is represented by the
standard sirens, i.e., the gravitational wave signal from
an inspiraling binary system to determine the absolute value of its
luminosity distances. Such original proposal, especially focusing on
inspiraling binary black holes (BH) and neutron stars (NS) can be
traced back to the paper of \cite{schutz1986determining}. The
breakthrough took place with the first direct detection of the
gravitational wave (GW) source GW150914
\citep{abbott2016observation}, as well as GW170817
\citep{abbott2017gw170817} with an electromagnetic counterpart,
which has opened an era of gravitational wave astronomy and added a
new dimension to the multi-messenger astrophysics. Since then,
extensive efforts have been made to use simulated GW data to
constrain cosmological parameters, which showed that the constraint
ability of GWs is comparable or better than the traditional probes,
if hundreds of GW events have been observed
\citep{cai2017estimating}. Compared with the observations of SNe Ia
in the EM domain, the self-calibrating GW signals could provide the
effective information of luminosity distances, independent of any
other distance ladders. More importantly, the greatest advantages of
GW lies in its ability to propagate freely through a perfect fluid
without any absorption and dissipation
\citep{ehlers1987propagation,ehlers1996wkb,weinberg1972gravitation},
in the Friedman-Lema\^{\i}tre-Robertson-Walker metric. Therefore,
when confronting the luminosity distance derived from SNe Ia with
that directly measured from GW sources, we may naturally propose a
scheme to investigate the opacity of the Universe, given the wealth
of current and future available data in gravitational wave (GW) and
electromagnetic (EM) domain. If the universe is opaque, the flux
from SN Ia received by the observer will be reduced, and we may
characterize this effect with a factor $e^{-\tau(z)}$, where
$\tau(z)$ is the optical depth related to the cosmic absorption. As
is discussed above, since the GWs travel in the Universe without any
absorption and scattering with dust, the observed luminosity
distance from SNe Ia is related to the true luminosity distance from
GW as
\begin{equation}
D_{L,\rm{SN}}=D_{L,\rm{GW}}e^{\tau(z)/2},
\end{equation}
More specifically, we will consider the simulated data of
gravitational waves from the third-generation gravitational wave
detector (the Einstein Telescope, ET), as well as the newly-compiled
type Ia supernovae (SNe Ia) data from Joint Light-curve Analysis
(JLA) sample and the Pantheon sample, in order to, compare
opacity-free distance from GW data and opacity-dependent distance
from SNe Ia.

This paper is organized as follows. The simulated GW data and the
current SNe Ia sample used in our work are presented in Section
\ref{data}. Section \ref{results} investigates the constraints these
data put on two different parameterizations of cosmic opacity.
Finally, the conclusions and discussions are presented in Section
\ref{conclusion}.

\begin{figure}
\centering
\includegraphics[width=1.1\hsize]{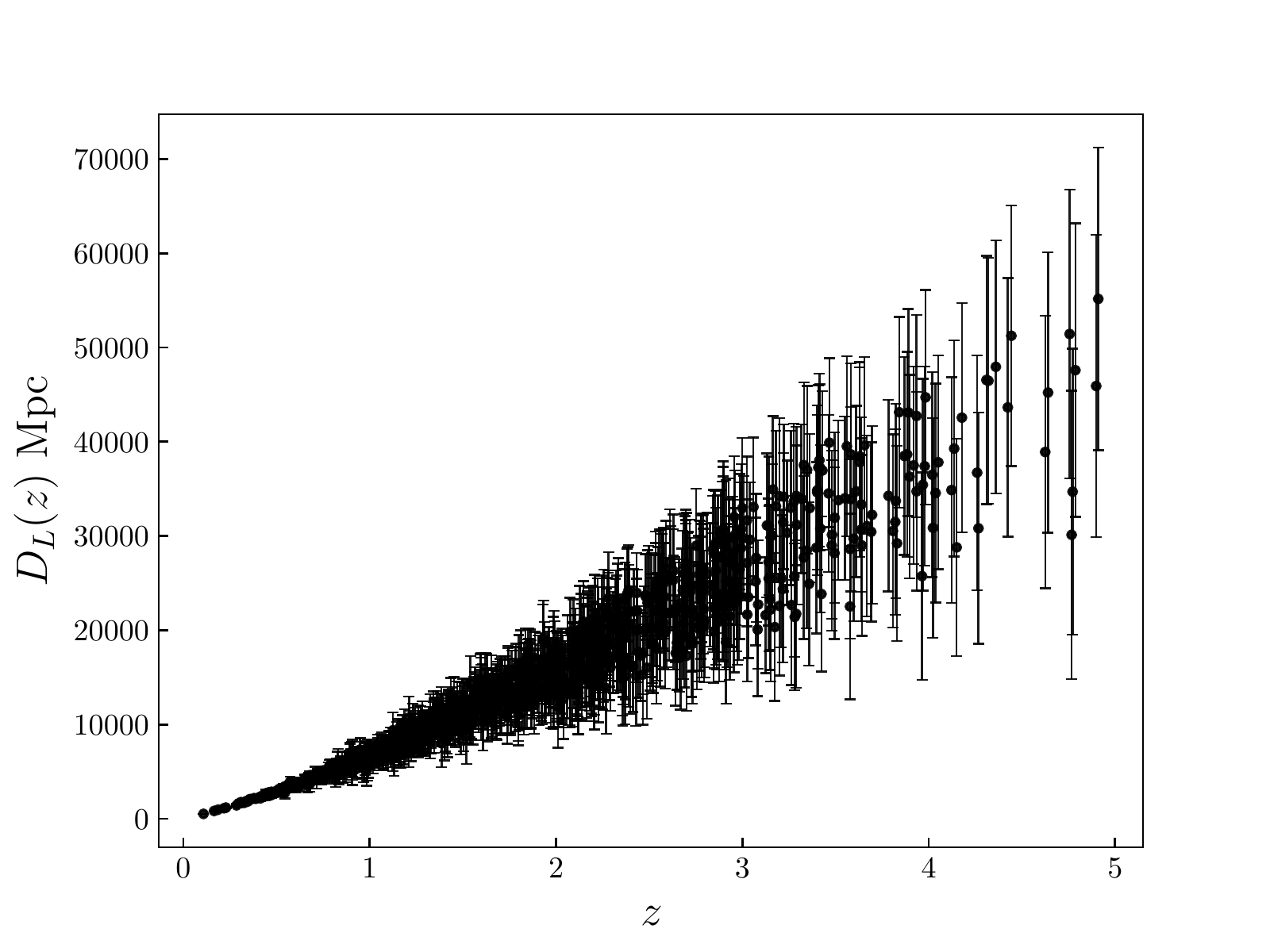}
\caption{The luminosity distance measurements from 1000 observed GW
events.}\label{ET_DL}
\end{figure}

\section{DATA} \label{data}

\subsection{Gravitational waves detected by ET}

First of all, we will briefly introduce the simulated observations
of GWs from the third generation of the ground-based GW detector,
Einstein Telescope (ET) \footnote{The Einstein Telescope Project,
\url{https://www.et-gw.eu/et/}}, which would be ten times more
sensitive than current advanced ground-based detectors covering the
frequency range of $1-10^4$ Hz. Theoretically, ET could detect GW
signals up to redshift $z\sim 2$ for the neutron star-neutron star
(NS-NS) mergers and $z\sim 5$ for black hole-neutron star (BH-NS)
mergers systems \citep{cai2017estimating}. These two GW sources are
of concern to our investigation in this paper, as the
electromagnetic (EM) signals are emitted during the merger
processes, allowing us to determine the redshift of sources.
Following the prediction in Ref.~\cite{cai2017estimating}, in the
framework of ET configurations, about $10^2$ GW sources with EM
signals will be detected per year. Let us briefly describe how we
simulate the GW sample.

I. The GW detectors based on the ET could measure the strain $h(t)$,
which quantifies the change of difference of two optical paths caused
by the passing of GWs. It can be expressed as the linear combination of
the two polarization states
\begin{equation}
h(t)=F_+(\theta, \phi, \psi)h_+(t)+F_\times(\theta, \phi, \psi)h_\times(t),
\end{equation}
where $F_{+,\times}$ are the beam pattern functions, $\psi$ denotes
the polarization angle, and ($\theta, \phi$) are the angles
describing the location of the source relative to the detector. The
explicit expressions of the antenna pattern functions for ET are
given by \citep{zhao2011determination}
\begin{align}
F_+^{(1)}(\theta, \phi, \psi)=&~~\frac{{\sqrt 3 }}{2}[\frac{1}{2}(1 + {\cos ^2}(\theta ))\cos (2\phi )\cos (2\psi ) \nonumber\\
                              &~~- \cos (\theta )\sin (2\phi )\sin (2\psi )],\nonumber\\
F_\times^{(1)}(\theta, \phi, \psi)=&~~\frac{{\sqrt 3 }}{2}[\frac{1}{2}(1 + {\cos ^2}(\theta ))\cos (2\phi )\sin (2\psi ) \nonumber\\
                              &~~+ \cos (\theta )\sin (2\phi )\cos (2\psi )].
\label{equa:F}
\end{align}
Because the three interferometers of the ET are arranged in an
equilateral triangle, the other two interferometer's antenna pattern
functions can also be obtained as $F_{+,\times}^{(2)}(\theta, \phi,
\psi)=F_{+,\times}^{(1)}(\theta, \phi+2\pi/3, \psi)$ and
$F_{+,\times}^{(3)}(\theta, \phi, \psi)=F_{+,\times}^{(1)}(\theta,
\phi+4\pi/3, \psi)$.

II. Focusing on the GW signals from the merger of binary systems
with component masses $m_1$ and $m_2$ (the corresponding total mass
is $M=m_1+m_2$, and the symmetric mass ratio is $\eta=m_1m_2/M^2$),
we can define the chirp mass $\mathcal{M}_c=M \eta^{3/5}$ and its
corresponding observational counterpart as $\mathcal{M}_{c,\rm
obs}=(1+z)\mathcal{M}_{c,\rm phys}$. For convenience,
$\mathcal{M}_c$ represents the observed chirp mass in the subsequent
analysis. If the change of orbital frequency over a single period is
negligible, by applying a stationary phase approximation, the
Fourier transform $\mathcal{H}(f)$ of the time domain waveform
$h(t)$ can be computed as
\begin{align}
\mathcal{H}(f)=\mathcal{A}f^{-7/6}\exp[i(2\pi ft_0-\pi/4+2\psi(f/2)-\varphi_{(2.0)})],
\label{equa:hf}
\end{align}
where $t_0$ is the epoch of the merger, while the definitions of the
functions $\psi$ and $\varphi_{(2.0)}$ can be found in
\citet{zhao2011determination}. Moreover, the Fourier amplitude
$\mathcal{A}$ is given by
\begin{align}
\mathcal{A}=&~~\frac{1}{D_L}\sqrt{F_+^2(1+\cos^2(\iota))^2+4F_\times^2\cos^2(\iota)}\nonumber\\
            &~~\times \sqrt{5\pi/96}\pi^{-7/6}\mathcal{M}_c^{5/6},
\label{equa:A}
\end{align}
where $\iota$ represents the angle of inclination of the binary's
orbital angular momentum, and $D_L$ is the luminosity distance that
can be derived from GW signals. Since the GW amplitude depends on
the so-called chirp mass and the luminosity distance, and the chirp
mass can be measured from the GW signal's phasing, we can extract
luminosity distance from the amplitude. Note that the GW sources
used in this work are caused by binary merger of a neutron star with
either a neutron star or black hole, which can generate an intense
burst of $\gamma$-rays (SGRB) with measurable redshift. More
importantly, from observational point of view, the SGRB is emitted
in a narrow cone, which indicates that one specific gravitational
wave event should be detected within the total beaming angle (e.g.,
$\iota<20^\circ$) \citep{nakar2007short}. In the following
simulations, we adopt the flat $\Lambda$CDM with $H_0=67.8
 \rm{km~s^{-1}Mpc^{-1}}$ and $\Omega_m=0.308$ as the fiducial
cosmological model, following the the most recent Planck results
\citep{ade2016planck}.

Given a waveform of GW, the signal-to-noise ratio (SNR) of a
detector can be written as
\begin{equation}
\rho=\sqrt{\left\langle \mathcal{H},\mathcal{H}\right\rangle},
\label{euqa:rho}
\end{equation}
where the inner product is defined as
\begin{equation}
\left\langle{a,b}\right\rangle=4\int_{f_{\rm lower}}^{f_{\rm upper}}\frac{\tilde a(f)\tilde b^\ast(f)+\tilde a^\ast(f)\tilde b(f)}{2}\frac{df}{S_h(f)},
\label{euqa:product}
\end{equation}
and $S_h(f)$ represents the one-side noise power spectral density
(PSD) characterizing the performance of a GW detector
\citep{zhao2011determination}. The lower cutoff frequency $f_{\rm
lower}$ is fixed to be 1 Hz, while the upper cutoff frequency,
$f_{\rm upper}$, is decided by the last stable orbit (LSO), $f_{\rm
upper}=2f_{\rm LSO}$, where $f_{\rm LSO}=1/(6^{3/2}2\pi M_{\rm
obs})$ is the orbit frequency at the LSO. For the network of three
independent interferometers in ET, the combined SNR can be
calculated as
\begin{equation}
\rho=\sqrt{\sum _ { i = 1 } ^ { 3 } \rho^2_i}
\end{equation}

III. According the Fisher information matrix, the instrumental
uncertainty of the measurement of $D_{L,GW}$ can be estimated. In
this analysis, we take the simplified case where the binary's
orbital plane is nearly face on and then the Fourier amplitude
$\mathcal{A}$ will be independent of the polarization angle $\psi$.
From theoretical point of view, the distance of source $D_L$ is
correlated with other parameters, especially the inclination angle
with possible values of $\iota=[0, 180^\circ]$. However, recent
analysis \citep{li2015extracting,cai2017estimating} indicated that
SGRBs, the electromagnetic counterparts of GWs, are likely to be
strongly beamed phenomena, which allow one to constrain the
inclination of the compact binary system and furthermore breaking
the distance-inclination degeneracy. More specifically, averaging
the Fisher matrix over the inclination $\iota$ with the limit
$\iota<20^\circ$ is approximately equivalent to taking $\iota=0$.
Therefore, we suppose that the luminosity distance $D_L$ is
independent of other GW parameters, and double its uncertainty
calculated from the Fisher matrix as the upper limit of the
instrumental error
\begin{align}
\sigma_{D_{L,GW}}^{\rm inst}\simeq \frac{2D_L}{\rho}.
\label{sigmainst}
\end{align}
Moreover, following the uncertainty budget described by
\citet{cai2017estimating}, the lensing uncertainty caused by the
weak lensing is modeled as $\sigma_{D_L}^{lens}/D_L=0.05z$.
Therefore, the distance precision per GW is taken as
\begin{align}
\sigma_{D_{L,GW}}&~~=\sqrt{(\sigma_{D_{L,GW}}^{\rm inst})^2+(\sigma_{D_{L,GW}}^{\rm lens})^2} \nonumber\\
            &~~=\sqrt{\left(\frac{2D_{L,GW}}{\rho}\right)^2+(0.05z D_{L,GW})^2}.
\label{sigmadl}
\end{align}
Let us clarify some simplified assumptions underlying our error
strategy listed above. In this paper, we consider only instrumental
and lensing uncertainties to derive the information of GW luminosity
distances. Note that the precise measurement of the chirp mass and
the redshift could constitute the biggest challenge of using GWs as
the standard sirens. On the one hand, as can be clearly seen from
Eq.~(5), the uncertainty related to the measurement of chirp mass
will contributes to the scatter of luminosity distances at high
redshifts and might reveal as a systematic effect at low redshifts.
On the other hand, with present sensitivity (of the advanced LIGO
and Virgo detectors), the localization accuracy is far from accurate
enough to identify the host galaxy and provide accurate measurement
of redshift. Therefore, the redshift inferred at the current
observational level will adds additional uncertainty to our
cosmological constraints. However, in the framework of the Einstein
Telescope, the third-generation detector with higher sensitivity,
one could expect the chirp mass to be accurately measured from the
GW signal's phasing, while the host galaxy can be identified from
the electromagnetic counterpart of GW (such as SGRB), the redshift
of which can be determined accurately by the follow-up observations.
Specially, following the recent analysis given the observations of
host galaxies, the peculiar velocity is typically set at 150-250
km/s and the corresponding redshift uncertainty is estimated to be
$\Delta z=0.001$ \citep{chen2018a}. Therefore, in our approach this
the redshift determination does not significantly contribute to the
scatter in the simulation results.

IV. We have simulated many catalogues of NS-NS and BH-NS systems,
with the masses of NS and BH sampled by uniform distribution in the
intervals [1,2] $M_{\odot}$ and [3,10] $M_{\odot}$. The ratio of the
possibility to detect the BHNS and BNS events is set to be $\sim
0.03$ \citep{cai2017estimating}. The sky position of GW source is
sampled from the uniform distribution on 2-dimensional sphere
\citep{lin2018testing}. In addition, the signal is identified as a
GW event only if the ET interferometers have a network SNR of
$\rho>8.0$, the SNR threshold currently used by LIGO/Virgo network
\citep{lin2018testing}. Finally, the redshift distribution of these
GW sources are taken as \citep{sathyaprakash2010cosmography}
\begin{equation}
P(z)\propto \frac{4\pi d_C^2(z)R(z)}{H(z)(1+z)}, \label{equa:pz}
\end{equation}
where $H(z)$ is the Hubble parameter of the fiducial $\Lambda$CDM,
$d_C=\int_0^z1/H(z)dz$ is co-moving distance, and $R(z)$ represents
the time evolution of the burst rate (see
\citep{schneider2001low,cutler2009ultrahigh} for details). Denoting
with $D_{L,GW}(z)$ the predicted value from our fiducial
cosmological model, we then assign to each GW, an opacity-free
luminosity distance randomly generated from a Gaussian distribution
centered on $D_{L,GW}(z)$ and $\sigma_{D_{L,GW}}$ from
Eq.~(\ref{sigmadl}). The simulated 1000 GW samples are shown in
Fig.~\ref{ET_DL}.

\begin{figure*}
\centering
\includegraphics[width=0.45\hsize]{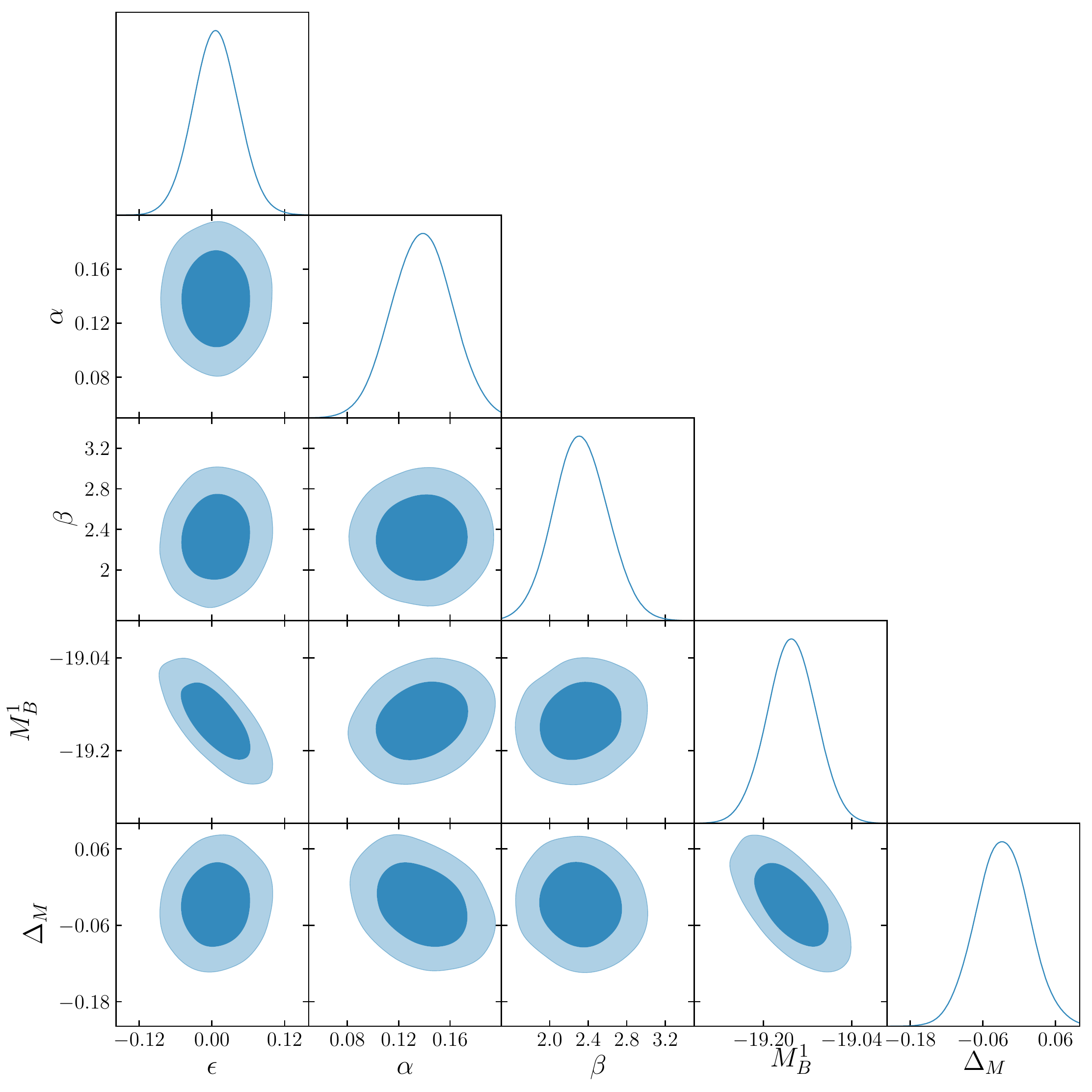} \includegraphics[width=0.45\hsize]{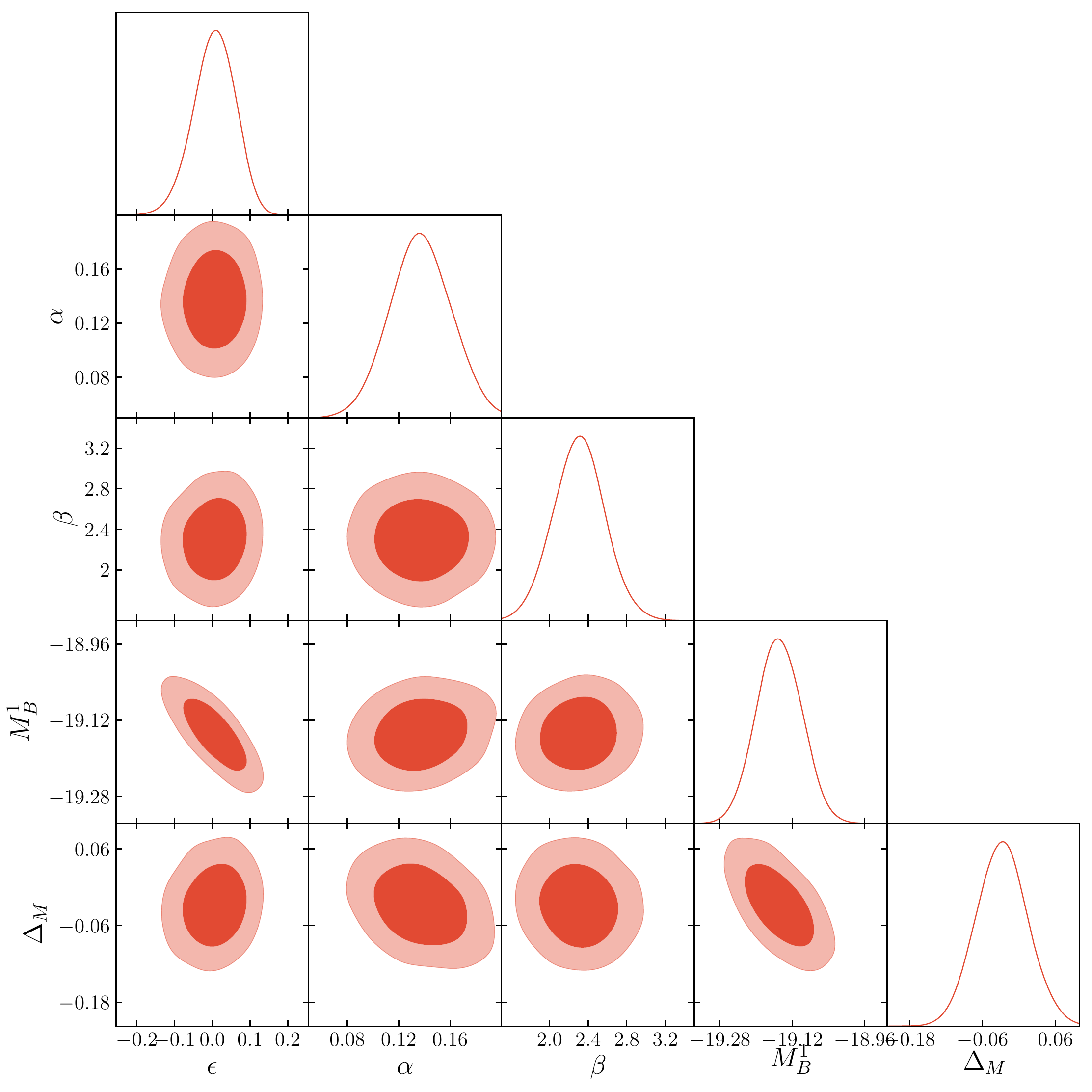}
\caption{The one-dimensional and two-dimensional distributions of
cosmic opacity parameter $\epsilon$ and SNe Ia nuisance parameters
($\alpha$, $\beta$, $M_B^1$ and $\Delta_M$) constrained from the JLA
sample in the P1 (left) and P2 (right) model, respectively.
}\label{JLA_P1}
\end{figure*}

\begin{figure*}
\centering
\includegraphics[width=0.45\hsize]{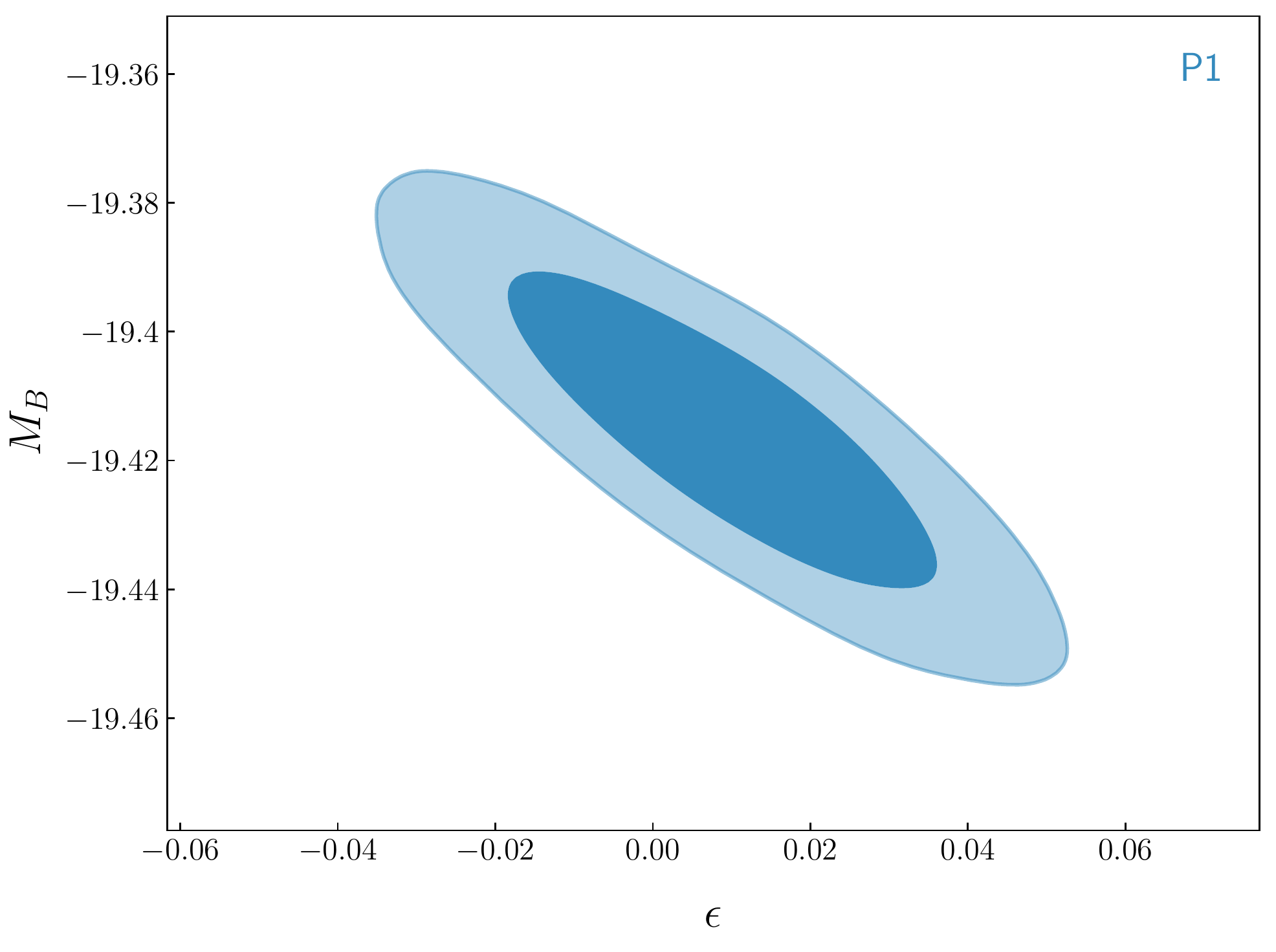}
\includegraphics[width=0.45\hsize]{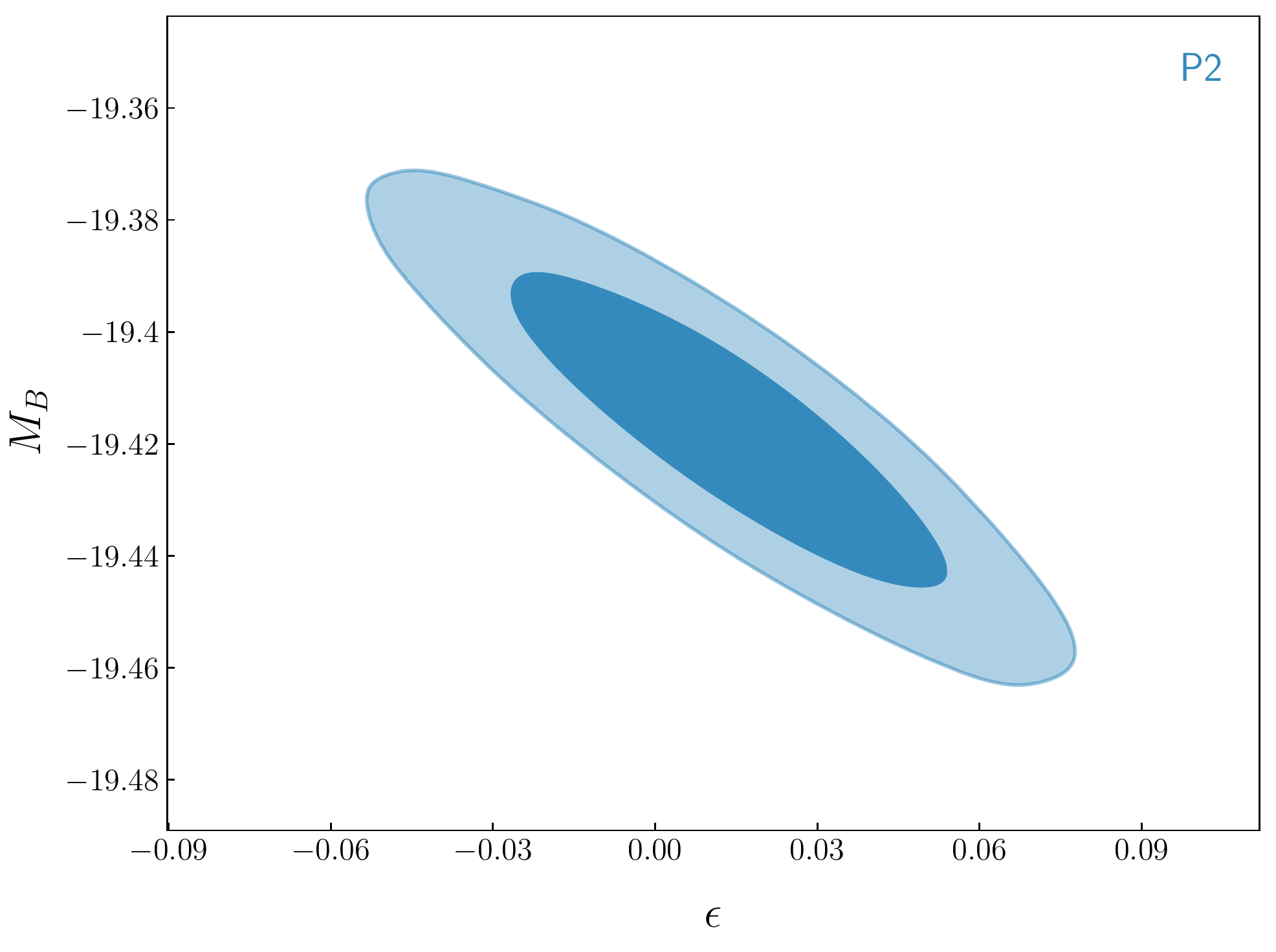}
\caption{The two-dimensional distributions of cosmic opacity
parameter $\epsilon$ and SNe Ia nuisance parameters ($M_B^1$)
constrained from the Pantheon sample in the P1 (left) and P2 (right)
model, respectively.}\label{pantheon}
\end{figure*}

\begin{table*}
\centering \setlength{\tabcolsep}{7mm}{
\begin{tabular}{cccccc}
\hline \hline
\multicolumn{6}{c}{JLA+ET}\\
\hline
&$\epsilon$&$\alpha$&$\beta$&$M_B^1$&$\Delta_M$\\
\hline
P1& $0.007\pm0.038$& $0.138\pm0.023$& $2.33\pm0.28$& $-19.149\pm0.044$& $-0.028\pm0.043$ \\

P2& $0.005\pm0.055$& $0.137\pm0.023$& $2.31\pm0.27$& $-19.149\pm0.050$& $-0.028\pm0.042$ \\

\hline\hline
\multicolumn{6}{c}{Pantheon+ET}\\
\hline
P1& $0.009\pm0.018$& \multicolumn{4}{c}{$M_B=-19.415\pm0.016$}\\
P2& $0.013\pm0.027$& \multicolumn{4}{c}{$M_B=-19.417\pm0.019$}\\
\hline \hline
\end{tabular}}
\caption{ Best-fit values with 1$\sigma$ standard error for the
cosmic opacity $\epsilon$ and SNe Ia nuisance parameters. }
\label{table1}
\end{table*}


\subsection{Latest Supernovae Ia observations}

Concerning the opacity-dependent distance modulus, we will turn to
the joint light-curve analysis (JLA) sample with 740 SNe Ia data
compiled by the SDSS-\uppercase\expandafter{\romannumeral2} and SNLS
collaborations \citep{betoule2014improved}, as well as the Pantheon
sample consisting of 1048 SN Ia recently released by Pan-STARRS1
(PS1) Medium Deep Survey \citep{scolnic2018complete}. \footnote{Note
that similar investigation of cosmic opacity from SNe Ia + GW was
given in \citet{wei2019}, which we received while working on this
paper.}

SNe Ia are used as ``standard candles'' to provide the most direct
indication of the accelerated expansion of the universe. The recent
discovery of a new gravitationally lensed SNe Ia from the
intermediate Palomar Transient Factory (iPTF) has also opened up a
wide range of possibilities of using strong lensing systems in
cosmology and astrophysics \citep{goobar2017iptf16geu,cao2018apj}.
Apart from developing a precise model able to determine the
standardization parameters directly from the physics of the SNe Ia
thermonuclear explosions, the only way to evaluate these parameters
is through the Hubble diagram. Indeed, one can express the distance
modulus of each SN Ia as a difference between its apparent and
absolute magnitude. For the JLA sample, the observed distance
modulus is
\begin{eqnarray}
\mu _ { \mathrm { SN } }&=& m_{B}^{*}+\alpha \cdot X _ { 1 } - \beta
\cdot \mathcal{C} - M _ { B },
\end{eqnarray}
where $m_{B}^{*}$ is the rest frame \textit{B}-band peak magnitude,
$X_1$ and $\mathcal{C}$ describe the time stretch of light curve and
the supernova color at maximum brightness, respectively. Moreover,
the parameter $M_B$ is the absolute \textit{B}-band magnitude, whose
value is determined by the host stellar mass $M_{\rm{stellar}}$ by a
step function
\begin{eqnarray}
M _ { B } = \left\{ \begin{array} { l l } { M _ { B } ^ { 1 } } & {
\text { for } M _ { \text { stellar } } < 10 ^ { 10 } M _ { \odot }
} \\ { M _ { B } ^ { 1 } + \Delta _ { M } } & { \text { otherwise. }
} \end{array} \right.
\end{eqnarray}
Thus, there are four nuisance parameters ($\alpha, \beta, M _ { B }
^ { 1 }$ and $\Delta _ { M }$) to be fitted, along with the
parameters characterizing the opacity of the universe. Recently, the
Pan-STARRS1 (PS1) Medium Deep Survey has released their Pantheon
compilation which consists of 1048 SNe Ia, which have been
extensively used to constrain cosmological models in
\citet{park2018observational,Huillier2018,qi2018what}. For the
Pantheon sample, the stretch-luminosity parameter $\alpha$ and the
color-luminosity parameter $\beta$ should be set to zero, and the
observed distance modulus is simply reduced to $\mu_{\rm{SN}}=
m_B^*-M_B$ \citep{scolnic2018complete}.

It should be noted that the distance modulus of the compiled SNe Ia
could provide the opacity-dependent luminosity distance as
$D_{L,SN}(z)=10^{\mu(z)/5-5}$. If the observed luminosity distance
from SNe Ia is related to the true luminosity distance from GWs by
$D_{L,\rm{SN}}^2=D_{L,\rm{GW}}^2e^{\tau(z)}$, the theoretical
distance modulus of a SNe Ia can be obtained as
\begin{equation}
\mu_{\rm{th}}(z)=5\log D_{L,\rm{GW}}+25+2.5(\log_{10}e)\tau(z).
\end{equation}
For a given SNe Ia data point, theoretically, we should select an
associated GW data point at the same redshift. In order to avoid any
bias of redshift differences between SNe Ia and GW, we adopt a
selection criterion that bins $D_{L,\rm{GW}}$ measurements within
the redshift range $\Delta z=|z_{\rm{SN}}-z_{\rm{GW}}|\leq0.005$.
One should note that the redshifts of observations are not
determined with infinite accuracy, which indicates that it is
unrealistic to decrease $\Delta z$ below the total 1$\sigma$ error
of observational redshifts
$\sigma_{z,tot}=\sigma_{z,SN}+\sigma_{z,GW}$ \citep{cao2011testing}.
For the observations of SNe Ia, the uncertainty of peculiar velocity
is set at the level of 300-400 km/s \citep{Wood07} and the
corresponding redshift uncertainty is $\sigma_{z,\rm SN}=0.001$
\citep{hicken2009improved}. For the observations of GW host
galaxies, the three-dimensional rms velocity (150-250 km/s)
corresponds to the redshift uncertainty of $\sigma_{z,\rm GW}=0.001$
\citep{chen2018a}. Therefore, in principle, $\Delta z=\sigma_{z,\rm
tot}=0.002$ should be considered in our work. However, considering
the observational difficulties in precisely identifying the host
galaxy and measuring GW redshift, it is not appropriate to use a
smaller window constraint. Thus we increase such uncertainty by a
factor and 2 (as the upper limit) and choose the SNe Ia points which
have the minimum acceptable redshift difference of the GW sample
$\Delta z\leq0.005$. Such selection criterion has been widely used
in the recent works \citep{Yang2017}, which tested the
potentialities of future GW sources to impose limit on possible
departures of the distance-duality relation with current strong
lensing observations \citep{cao2012constraints,cao2015cosmology}.

The likelihood estimator is determined by $\chi^2$ statistics
\begin{equation}
\chi^2=(\mu_{th}-\mu_{\rm{SN}})\cdot \textbf{Cov}^{-1} \cdot
(\mu_{th}-\mu_{\rm{SN}}),
\end{equation}
where $\textbf{Cov}$ is the covariance matrix. For robustness and
simplicity, we only consider the statistical uncertainty, and it
defined by
\begin{equation}
\textbf{Cov}=\textbf{D}_{\rm{stat}}+\sigma^2_{\rm{GW}},
\end{equation}
where $\sigma_{\rm{GW}}$ is the uncertainty of $D_{L,\rm{GW}}$, and
$\textbf{D}_{\rm{stat}}$ is the diagonal part of the statistical
uncertainty, whose expression is
\begin{eqnarray}
\left( \textbf{D}_{\rm{stat}}\right) _ { i i } =&& \sigma _ { m _ { B } , i } ^ { 2 } + \alpha ^ { 2 } \sigma _ { X _ { 1 } , i } ^ { 2 } + \beta ^ { 2 } \sigma _ { \mathcal { C } , i } ^ { 2 }+ 2 \alpha C _ { m _ { B }  X _ {1} , i }\nonumber \\
&& - 2 \beta C _ { m _ { B } \mathcal { C } , i} - 2 \alpha \beta C
_ { X _ { 1 } \mathcal { C } , i },
\end{eqnarray}
where $\sigma _ { m _ { B } , i }$ , $\sigma _ { X _ { 1 } , i }$ ,
and $\sigma _ { \mathcal { C } , i }$ denote the errors of the peak
magnitude and light curve parameters of the $i$th SN Ia.
$C_{m_{B}X_{1},i}$, $C_{m_{B}\mathcal{C},i}$ and $C _ { X _ { 1 }
\mathcal{C} , i }$ represent the covariances of $m_B$, $X_1$,
$\mathcal{C}$ for the $i$th SN Ia. For the Pantheon sample, however,
the stretch-luminosity parameter $\alpha$ and the color-luminosity
parameter $\beta$ are set to zero, whose statistical uncertainty
simplifies to $\textbf{D}_{\rm{stat}}=\sigma^2 _ { m _ { B }}$.

In this work, we directly adopt the observational quantities
($m_{B}^{*}, X_{1}, \mathcal{C}$) from the JLA sample and Pantheon
sample to constrain the cosmic opacity ($\tau(z)$). By marginalizing
the nuisance parameters ($\alpha$, $\beta$,  $M_{B}^{1}$ ($M_B$),
$\Delta_{M}$), one can obtain a cosmology-independent constraint on
the opacity and justify whether the cosmic opacity has a dependence
on the nuisance parameters. The constraints on the parameter are
derived by evaluating the likelihood distribution function,
$L\propto exp(-\chi^2/2)$, with the corresponding $\chi^2$ defined
in Eq.~(15). We choose to determine the best-fit values and the
marginalized errors of each model parameter through the Markov chain
Monte Carlo (MCMC) method, which has been extensively applied in
cosmological studies
\citep{cao2015cosmology,li2016comparison,Qi2017,Xu2018}. The
advantage of the MCMC method is that it allows for a simple
inclusion of priors and a comprehensive study of the effects of
systematic uncertainties. Our code is based on the publicly
available emcee Python module \citep{emcee}.

\section{ Cosmic opacity parameterizations and constraints} \label{results}

Regarding the parametrization of the opacity of the Universe, a
model-independent test has been extensively discussed in the above
quoted papers \citep{li2013cosmic,liao2013testing,liao2015universe}.
In general, $\tau$ can be treated as parameterized functions of the
redshift,
\begin{eqnarray}
P1. \  \tau(z)& = & 2\epsilon z, \\ \nonumber P2. \   \tau(z)&= &
(1+z)^{2\epsilon}-1. \\ \nonumber
\end{eqnarray}
which are not strongly wavelength dependent on the optical band. The
former linear parametrization is inspired on similar expressions for
DDR, which can be derived from the parameterization
$D_L(z)=D_A(z)(1+z)^{2+\epsilon}$ for small $\epsilon$ and redshift.
The latter parametrization, which is basically similar to the former
one for $z\ll 1$ but could differ when $z$ is not very small. For
the two models, one should expect the likelihood of $\epsilon$ to
peak at $\epsilon=0$, if it is consistent with photon conservation
and there is no visible violation of the transparency of the
Universe. The graphic representations of the probability
distribution of the opacity parameter are presented in Fig.~\ref{JLA_P1} and \ref{pantheon},
respectively. We give the 1-D distributions for each parameter
($\epsilon$; $\alpha$, $\beta$, $M_{B}^{1}$ ($M_B$), $\Delta_{M}$)
and 1$\sigma$, 2$\sigma$ contours for the joint distributions of any
two parameters. The corresponding best-fit parameters are summarized
in Table~\ref{table1}, along with the 1$\sigma$ standard deviations for each.

As one may see, the analyses are consistent with zero cosmic opacity
within 68.3\% confidence level for both of SNe Ia samples, implying
that there is no significant deviation from the transparency of the
Universe at the current observational data level. Similar to the
results obtained by examining the cosmic opacity in a particularly
low redshift range ($z<0.890$) \citep{li2013cosmic}, we find that an
almost transparent universe is also favored by the JLA sample at
higher redshifts ($z<1.30$): the best-fit $\epsilon$ parameter with
$1\sigma$ confidence level is $0.007\pm0.038$ for P1 function.
Therefore, the upper limit for the optical depth related to the
cosmic absorption per $\rm{Mpc}$ is about $10^{-5}~\rm{Mpc}^{-1}$ at
$68.3\%$ C.L. More interestingly, we find that our constraints on
the nuisance parameters (see Table~\ref{table1}) are very different
from those results of \citet{betoule2014improved}:
$\alpha=0.140\pm0.006$, $\beta=3.101\pm0.072$ and
$M_B=-19.04\pm0.01$, which are derived from a fit to the flat
$\Lambda$CDM cosmology. Therefore, the consideration of cosmic
opacity might effectively affect the values of SNe Ia nuisance
parameters, which can be particularly seen from the constraints in
$(\epsilon, ~ M_B^{1})$ plane. We still find strong degeneracies
between $\epsilon$ and $M_B^{1}$, i.e., a lower absolute
\textit{B}-band magnitude of SNe Ia will lead to a larger value of
the cosmic opacity, which not only attests to the reliability of our
calculation, but also confirms that the cosmic opacity parameter is
not independent of the nuisance parameters. Working on the Pantheon
sample, one can clearly see that the currently larger data improves
the constraints on model parameters significantly. From the above
results, the parameter $\epsilon$ capturing the transparency of the
Universe seems to be vanishing: $\epsilon=0.009\pm0.018$ for P1
function. Compared with the JLA SNe Ia standard candles, the
advantage of the Pantheon sample is that SNe Ia are observed at much
higher redshifts ($z\sim 2.26$), which motivate us to investigate
the cosmic opacity in the early universe. The strong degeneracies
between the cosmic opacity parameter $\epsilon$ and the intrinsic
brightness parameter $M_B$ is also illustrated in
Fig.~\ref{pantheon}.

\begin{table}
\centering
\begin{tabular}{lcc}
\hline \hline
Data & $\epsilon$ (P1) &$\epsilon$ (P2) \\
\hline JLA + ET & $0.007\pm0.038$& $0.005\pm0.055$ \\
Pantheon + ET & $0.009\pm0.018$& $0.013\pm0.027$ \\
\hline
Union2.1 + Cluster\citep{li2013cosmic}&  $0.009^{+0.059}_{-0.055}$& $0.014^{+0.071}_{-0.069}$ \\
Union2.1 + $H(z)$ \citep{liao2013testing} & $-0.01_{-0.10}^{+0.10}$ & $-0.01_{-0.12}^{+0.12}$ \\
JLA + $H(z)$ \citep{liao2015universe}& $0.07^{+0.107}_{-0.121}$ & $\Box$ \\

\hline \hline
\end{tabular}
\caption{Summary of the best-fit value for the cosmic opacity
parameter obtained from different observations.}
\label{comparsion_results}
\end{table}

It is worth investigating how the constraints depend on the assumed
$\tau(z)$ parameterization. For the P2 parametrization, the results
derived from the JLA sample and Pantheon sample are shown in
Fig.~\ref{JLA_P1}-\ref{pantheon} and Table~\ref{table1}. The
best-fit cosmic-opacity parameters with $1\sigma$ confidence level
are $\epsilon=0.005\pm0.055$ and $\epsilon=0.013\pm0.027$,
respectively. Comparing the constraints on the two $\tau(z)$
parameterizations in Table \ref{table1}, we can see that the
dependence of test results on the above-chosen parameterizations for
$\tau(z)$ is relatively weak. Indeed, the 68\% confidence ranges are
well overlapped for the two $\tau(z)$ functions so that one could
draw conclusions on cosmic opacity in a roughly model independent
way. However, it should be noted that P2 function may be reconciled
with the data only if smaller $\epsilon$ values are used, i.e., a
smaller $\epsilon$ partially compensates for the different scalings
with $z$ of the two cases considered, which highlights the
importance of choosing a reliable parameterization for $\tau(z)$ in
order to better check the cosmic opacity validity at any redshift.

Now it is worthwhile to compare our forecast results with some
actual tests involving the angular diameter distances from various
astrophysical probes in the EM window. The recent determinations of
the cosmic-opacity parameters from different independent
cosmological observations are also listed in Table
\ref{comparsion_results}. \citet{li2013cosmic} combined two galaxy
cluster samples \citep{de2005measuring,bonamente2006determination}
with luminosity distances from the largest Union 2.1 type Ia
supernova sample. The analysis results show that an almost
transparent universe is favored by Filippis et al. sample but it is
only marginally accommodated by Bonomente et al. samples at 95.4\%
confidence level. Another analysis was also performed in
\citet{liao2013testing}, by fitting the luminosity distance of Union
2.1 SNe Ia with the newly published 28 observational Hubble
parameter data. The results, in the framework of three
model-independent methods (nearby SNe Ia method, interpolation
method and smoothing method), converged to a point that the effects
of cosmic opacity are vanished. Such methodology was recently
extended by \citet{liao2015universe}, who examined the residuals
between the constructed opacity-free luminosity distances from
$H(z)$ determinations and distance estimation in type Ia supernovae
observations with variable light-curve fitting parameters. A
transparent universe is currently consistent with the current EM
data. By comparing the results at 1$\sigma$ C. L., we obtain the
error bar 65\% smaller than that from \citep{liao2015universe}, when
the P1 parametrization is considered. By considering our results and
those from \citet{liao2013testing}, we obtain that our error bars
are 60\% and 55\% smaller when the P1 and P2 functions are
considered. Finally, focusing on the Pantheon compilation which
consists of more SNe Ia, one could expect much smaller error bars
when the P1 and P2 functions are considered, more precisely, 80\%
and 75\%, respectively \citep{liao2013testing}. Therefore, given the
wealth of future available data in both EM and GW domain
\citep{cao2017ultra,Caomulti,Qi2017,zheng2017ultra,cao2019milliarcsecond},
our results show that strong constraints on cosmic opacity (and
consequently on background physical mechanisms) can be obtained in a
cosmological-model-independent fashion.

\section{Conclusions and Discussions} \label{conclusion}

The first direct detection of the gravitational wave (GW) source
with an electromagnetic counterpart has opened an era of
gravitational wave astronomy and added a new dimension to the
multi-messenger astrophysics. Compared with the observations of SNe
Ia in the EM domain, the greatest advantages of GW signals lies in
its ability to propagate freely through a perfect fluid without any
absorption and dissipation in the FLRW metric. Therefore, one can be
confident that, future GW data will make it possible not only to
improve the precision of the constraints on cosmological models, but
also, test the cornerstones of observational cosmology. More
specifically, the cosmic opacity, the importance of which is usually
underrated, stands out as one of the fundamental pillars our
interpretation of astrophysical data.

In this paper, we propose a scheme to investigate the opacity of the
Universe in a cosmological-model-independent way, with the
combination of current and future available data in gravitational
wave (GW) and electromagnetic (EM) domain. More specifically, we
consider the simulated data of gravitational waves from the
third-generation gravitational wave detector (the Einstein
Telescope, ET), as well as the newly-compiled type Ia supernovae
(SNe Ia) data from Joint Light-curve Analysis (JLA) sample and the
Pantheon sample, in order to compare the opacity-free distance from GW
data and opacity-dependent distance from SNe Ia. Two redshift-
dependent parametric expressions: $\tau(z)=2\epsilon z$ and
$\tau(z)=(1+z)^{2\epsilon}-1$ are considered to describe the optical
depth associated with the cosmic absorption. Here we summarize our
main conclusions in more detail:

\begin{itemize}

\item We find that the optimized cosmic-opacity parameters change quantitatively, though
the qualitative results and conclusions remain the same, independent
of which kind of the sample is used, i.e., there is no significant
deviation from the transparency of the Universe at the current
observational data level. Similar to the previous results obtained
by examining the cosmic opacity in a particularly low redshift range
($z<0.890$), an almost transparent universe is strongly favored by
the JLA sample and the Pantheon sample at much higher redshifts
($z\sim 1.30$ and $z\sim 2.26$). However, we still find strong
degeneracies between the cosmic opacity parameter $\epsilon$ and the
intrinsic brightness parameter $M_B$, i.e., a lower absolute
\textit{B}-band magnitude of SNe Ia will lead to a larger value of
the cosmic opacity, which confirms that the cosmic opacity parameter
is not independent of the nuisance parameters. As a consequence,
this source of systematic error should be fully taken into account
with future data.

\item The results suggest that the tests of cosmic opacity are not significantly sensitive to
the parametrization for $\tau(z)$. Indeed, the 68\% confidence
ranges are well overlapped for the two $\tau(z)$ functions so that
one could draw conclusions on cosmic opacity in a roughly model
independent way. However, it should be noted that P2 function may be
reconciled with the data only if smaller $\epsilon$ values are used,
i.e., a smaller $\epsilon$ partially compensates for the different
scalings with $z$ of the two cases considered, which highlights the
importance of choosing a reliable parameterization for $\tau(z)$ in
order to better check the cosmic opacity validity at any redshift.

\item Comparing our forecast results with some actual tests involving the angular diameter distances from various
astrophysical probes in the EM window, we obtain that future
measurements of the luminosity distances of gravitational waves
sources will be much more competitive than the current analyses.
Therefore, given the wealth of more precise data, especially the GW
data in the coming years, we may expect more vigorous and convincing
constraints on the cosmic opacity (and consequently on background
physical mechanisms) and a deeper understanding the intrinsic
properties of type Ia supernovae in a cosmological-model-independent
way.

\end{itemize}

\section*{Acknowledgments}
This work was supported by National Key R\&D Program of China No.
2017YFA0402600; the National Natural Science Foundation of China
under Grants Nos. 11690023, 11503001, 11633001 and 11873001; Beijing
Talents Fund of Organization Department of Beijing Municipal
Committee of the CPC; the Fundamental Research Funds for the Central
Universities and Scientific Research Foundation of Beijing Normal
University; and the Opening Project of Key Laboratory of
Computational Astrophysics, National Astronomical Observatories,
Chinese Academy of Sciences. J.-Z. Qi was supported by the
Fundamental Research Funds for the Central Universities N180503014
and N182410008-1. Y. Pan was supported by the Scientific and
Technological Research Program of Chongqing Municipal Education
Commission(Grant No KJ1500414); Chongqing Municipal Science and
Technology Commission Fund(cstc2015jcyjA00044, cstc2018jcyjAX0192).

\bibliography{opacity}

\end{document}